\documentclass[12pt,preprint]{aastex}

\received{5 March 2002}
\revised{7 May 2002}
\accepted{7 May 2002}

\cpright{AAS}{2002}

\shortauthors{Hjorth et al.}
\shorttitle{Time delay of \rxj}

\slugcomment{Submitted to ApJL, 5 March 2002; accepted 7 May 2002}

\newcommand{\hst}{{\sl HST\/}}
\newcommand{\chandra}{{\sl Chandra X-ray Observatory\/}}
\newcommand{\xmm}{{\sl XMM-Newton\/}}

\newcommand{\kms}{{\rm km\,s}^{-1}}
\newcommand{\ho}{\kms\ {\rm Mpc}^{-1}}

\newcommand{\rxj}{\objectname{RX~J0911.4+0551}}
\newcommand{\rxja}{RX~J0911.4+0551}

\begin{document}

\title{The time delay of the quadruple quasar \rxj%
             \thanks{Based on observations made with the Nordic Optical 
	      Telescope, operated on the island of La Palma jointly by 
              Denmark, Finland, Iceland, Norway, and Sweden, in the 
              Spanish Observatorio del Roque de los Muchachos of the 
              Instituto de Astrofisica de Canarias.}}

\author{Jens~Hjorth\altaffilmark{2},         
Ingunn~Burud\altaffilmark{3,4},              
Andreas~O.~Jaunsen\altaffilmark{5},          
Paul~L.~Schechter\altaffilmark{6},           
Jean-Paul~Kneib\altaffilmark{7},             
Michael~I.~Andersen\altaffilmark{8},         
Heidi~Korhonen\altaffilmark{8},              
Jacob~W.~Clasen\altaffilmark{9},             
A.~Amanda~Kaas\altaffilmark{9},              
Roy~\O stensen\altaffilmark{10},             
Jaan~Pelt\altaffilmark{11},                  
and
Frank~P.~Pijpers\altaffilmark{12}            
}

\altaffiltext{2}{Astronomical Observatory, University of Copenhagen,
Juliane Maries Vej 30, DK--2100 Copenhagen~\O, Denmark; 
\email{jens@astro.ku.dk}}
\altaffiltext{3}{Institut d'Astrophysique et de G{\'e}ophysique de Li{\`e}ge,
Avenue de Cointe 5, 4000 Li{\`e}ge, Belgium}
\altaffiltext{4}{Space Telescope Science Institute, 3700 San Martin Drive,
Baltimore, MD 21218, USA}
\altaffiltext{5}{European Southern Observatory, Casilla 19001, Santiago 19,
Chile}
\altaffiltext{6}{Department of Physics, Massachusetts Institute of Technology, 
Cambridge, MA 02139}
\altaffiltext{7}{Observatoire Midi-Pyr\'en\'ees, 14 Av.~E. Belin, 
F--31400 Toulouse, France}
\altaffiltext{8}{Division of Astronomy, University of Oulu, P. O. Box 3000,
FIN--90014 Oulu, Finland} 
\altaffiltext{9}{Nordic Optical Telescope, Apartado 474, E--38700 
Santa~Cruz de La Palma, Canary Islands, Spain}
\altaffiltext{10}{Isaac Newton Group of Telescopes, E--38700 Santa Cruz de
La Palma, Canary Islands, Spain}
\altaffiltext{11}{Tartu Observatory, T\~{o}ravere, 61602, 
Estonia}
\altaffiltext{12}{Theoretical Astrophysics Center, University of Aarhus, 
DK--8000~\AA rhus C, Denmark}

\begin{abstract}

We present optical lightcurves of the gravitationally lensed components 
A~($\equiv$~A1+A2+A3) and B of the quadruple quasar \rxj\ ($z=2.80$). The 
observations were primarily obtained at the Nordic Optical Telescope 
between 1997 March and 2001 April and consist of 74 $I$-band data points 
for each component. The data allow the measurement of a time delay of
$146\pm8$ days (2$\sigma$) between A and B, with B as the leading component. 
This value is significantly shorter than that predicted from simple models 
and indicates a very large external shear.  Mass models including the main 
lens galaxy and the surrounding massive cluster of galaxies at $z=0.77$, 
responsible for the external shear, yield 
$H_0 = 71\pm4\ ({\rm random}, 2\sigma)\pm 8\ ({\rm systematic})\ \ho$.  
The systematic model uncertainty is governed by the surface-mass density
(convergence) at the location of the multiple images. 

\end{abstract}


\section{Introduction}

The light from a gravitationally lensed, multiply imaged quasar 
travels along slightly different paths towards the observer.  
This gives rise to a time delay: 
if the quasar is variable the same variation will be seen in the multiple 
images at different times. The time delay is inversely proportional to the 
Hubble constant, thus offering a means to measure its cosmological value in 
a direct way \citep{refsdal64}.  Depending on the source and lens redshifts 
the time delay will have a weak dependence on other cosmological 
parameters. The time delay also depends on the gravitational 
potential and hence the detailed mass distribution towards the quasar, thus 
complicating the measurement of cosmological parameters but providing 
powerful constraints on distant structures such as galaxies, groups, and 
clusters. 

With the aim of constraining cosmological parameters and the mass distribution 
in galaxies and clusters we have conducted a survey of time delays for 
multiply imaged (gravitationally lensed) quasars with the Nordic Optical 
Telescope (NOT).  A description of the 
program and a measurement of the time delay in our first target, the double 
quasar B~1600+434 which is lensed by an edge-on disk galaxy, was presented 
by \citet{burud00}. We here present the observations, data analysis and time 
delay of the second survey target, the quadruple quasar \rxj\ ($z=2.80$) which 
is lensed by a galaxy and a cluster at $z=0.77$. The time delay of the
third survey target, \objectname{SBS~1520+530}, is presented by 
\citet{burud02}.

\rxj\ was discovered by \citet{bade} in the ROSAT All Sky Survey.
\citet{burud98} found optical and near infrared evidence that \rxj\ is a 
quadruply lensed quasar with an unusual image configuration requiring a large 
external shear. The origin of this large external shear was attributed to 
a possible nearby cluster at a photometric redshift $z=0.7\pm0.1$.  
\citet{kneib00} confirmed the existence of a massive cluster and measured the 
redshift $z=0.769$ and velocity dispersion $\sigma=836^{+180}_{-200}~\kms$ of  
the cluster (to which the main lens galaxy belongs). From \chandra\ 
observations \citet{morgan01} found a cluster temperature of
$2.3^{+1.8}_{-0.8}$ keV and a 2--10 keV luminosity of
$7.6^{+0.6}_{-0.2} \times 10^{43}$ erg s$^{-1}$.

The  three  A  (A1,  A2,  A3)  components  of  \rxj\  are  very  close
($\theta_{\rm A1} - \theta_{\rm A2} = 0\farcs478$, 
$\theta_{\rm A2} - \theta_{\rm A3} = 0\farcs608$) and the time delays 
between  them are expected to be short, of the order of  less than one week.  
Therefore, we focused on determining the time  delay between  A and  B, 
separated by  3.1 arcsec,  expected to  be of  the order  of many 
months.  The images  are fairly bright with  mean $I$-band magnitudes of  
17.2 and  19.2 for A  and B respectively.   Early observations  showed  that 
the  QSO is  strongly variable with  a time  delay of about  200 days  
\citep{hjorth00}.  We here present an analysis of the full data set.

\section{Lightcurves}

$I$-band images were obtained at the NOT about 
once a week between September 1998 and April 2001. These regular monitoring 
data were supplemented with a few early data points obtained at NOT and 
MDM between March 1997 and June 1998. The target is below the horizon at 
the NOT in July and August. There were additional gaps in the lightcurves 
due to periods of bad weather and time allocated to Spanish and international 
observing programs. Three 
different instruments were used: ALFOSC (Andaluc\'\i a Faint  Object  
Spectrograph and Camera), HiRAC (High Resolution Adaptive Camera) and the 
stand-by camera StanCam, equipped with detectors yielding pixel scales of 
0\farcs189, 0\farcs107 and  0\farcs176 respectively. The $I$-band was chosen 
to minimize the sensitivity to lunar phase. One data point typically 
consisted of three exposures of 3--5 minutes each. The seeing  varied from 
0\farcs6 to 1\farcs5, with 0\farcs9 being the most frequent value. 
We typically obtained a signal-to-noise ratio of 200--300 for the 
summed A components and 50--100 for the B component.  

The data  were reduced and  analyzed as described  by \citet{burud00}.
Preprocessing was done by  dedicated pipelines while fringe correction
and cosmic-ray  removal were  performed manually. Data  from different
detectors  were brought  onto  the same  photometric reference  system
\citep{burud98} via appropriate color terms.  Three  reference  stars   
with  known magnitudes    were   used   to    calibrate   the    photometry   
(see Table~\ref{refstars} for their  coordinates and magnitudes). 
The photometry of the quasar images was performed by applying the MCS 
deconvolution algorithm \citep{Magain}. This algorithm has already been 
used to analyze the data of several blended lensed quasar images 
(e.g., \citet{burud98}, \citet{burud00}).  Its main advantage is its ability 
to use all, even rather poor, data, irrespective of image quality and 
lunar phase. The final deconvolved image is produced by simultaneously 
deconvolving the individual frames of the same object from all epochs. 
The positions of the quasar images and the shape of the lensing galaxy are 
the same for all the frames and are therefore constrained using the total 
S/N of the data-set. The intensity of the point sources are allowed 
to vary from image to image, hence producing the lightcurves.  Images 
of \rxj\ and  the resulting lightcurves  are presented in  Figs.~1 and 2.

\section{Time delay}

The  $I$-band lightcurves   (Fig.~\ref{lightcurve}) contain  74  data
points for each component. As predicted by  theory, the lightcurves 
show that B is the leading component.  A pronounced V-shaped feature at 
JD~2451300 (May/June~1999) is seen in the A component followed by 
several decreases and upturns. These are preceeded by similar  features  
in the  B  component which allows  the determination of a rough time 
delay of about 150~days. The data points are available upon request.

A quantitative  analysis of the light  curves  was performed using the
four methods  described  by  \citet{burud00}.    The SOLA method 
\citep{pijpers97} does not provide a definite time delay but is consistent
with the results of the minimum dispersion method \citep{pelt96} and the 
two novel methods introduced by \citet{burud01}.  The time-delay estimates 
and magnitude offsets  obtained  from the three different  methods
are consistent with each other and presented in Table~\ref{methods}. 

It is readily apparent that no simple time translation will turn the A 
curve into the B curve.  Thus, `external' variations in the time-delay
corrected flux ratios must be present.  This is confirmed  by our analysis 
of the individual A1, A2 and A3   lightcurves which exhibit similar overall 
trends but different detailed shapes. The time delays determined from 
the individual components are consistent with the expectation of small 
time delays between the A components and with the average B--A time delay 
(see Table~3).

The `external' variations, due e.g.\ to microlensing (see 
\citet{burud00}) are best taken into account by the methods
introduced by \citet{burud01}. In fact, the cause of the failure of 
the SOLA method lies in the nature of the external variations in 
\rxj, which are of bigher order than linear in time over the entire 
time series. This violates an assumption under which that method is 
derived. The best method for dealing with such high-order
variations is the iterative method.  We therefore adopt the time 
delay as determined by the iterative method (see Fig.~3) and conclude that 
the 
time delay between A and B in \rxj\ is $146\pm 8$~days ($2 \sigma$). 
With a $2 \sigma$ error of 5 \% this is among the most precise time 
delays determined for any lens system.

%
%

\section{Discussion}

In modeling the system we used a cosmology with $\Omega_m = 0.3$, 
$\Omega_\Lambda = 0.7$, and $H_0 = 100 h ~\kms~{\rm Mpc}^{-1}$.
Adopting an open Universe with $\Omega_m = 0.3$, $\Omega_\Lambda = 0$
increases the model time delay (and derived $H_0$) by 4.1 percent, whereas 
an Einstein--de Sitter Universe with $\Omega_m = 1.0$, $\Omega_\Lambda = 0$ 
leads to a decrease of 10.0 percent. 

The `yardstick' potential model of \cite{schechter00} involves a 
two-dimensional potential given by $\psi\sim \theta^{1+\alpha}$ with 
$\alpha=-0.8$ and an external shear with $\kappa=\gamma=0.307$. 
In the adopted cosmology, this model predicts a time delay of 
111$h^{-1}$ with an estimated uncertainty of $\pm22h^{-1}$ days due 
to the uncertainty in $\alpha$ from the scatter in the observed shapes 
of nearby elliptical galaxies. In the model presented by \cite{kneib00}
the predicted time delay is $112.5\pm17.5h^{-1}$, consistent with
the `yardstick' prediction of \cite{schechter00}, and with a slightly
smaller uncertainty.

We independently modeled the system as described by \cite{kneib00}.
The model includes the main lensing galaxy, the cluster of galaxies,
and individual galaxies in the cluster.  We used the measured velocity 
dispersion of \cite{kneib00} to constrain the cluster mass and adopted 
more realistic contraints on the ellipticity of the cluster mass distribution 
and a tighter mass-luminosity relation for the cluster galaxies.  The main 
model uncertainty concerns the value of the cluster convergence $\kappa$ at 
the location of the multiple images. A large $\kappa$ gives rise to a small 
predicted time delay and derived $H_0$. The value of $\kappa$ is governed by 
the mass of the cluster, its mass profile, and the effects of possible 
substructure in the cluster. We find that $0.20 < \kappa < 0.28$ is required 
for a good fit. This range in $\kappa$ translates into different values for 
the mass and velocity dispersion of the cluster, depending on the mass profile 
used and the ellipticity of the cluster mass distribution. In addition to 
the smooth cluster convergence there is a contribution to the total
convergence of about 0.06 from the individual galaxies in the cluster
(not including the main lens).

The refined model prediction of the (flux-weighted or straight) mean 
B--A time delay is $104\pm11h^{-1}$ (the individual time delays between 
the A components are less than 1.5 $h^{-1}$ days, generally in the sequence 
A2, A1, A3). Using the measured B--A time delay of $146\pm8$ days (2$\sigma$),
the resulting value of the Hubble constant is
$H_0 = 71\pm4\ ({\rm random}, 2\sigma)\pm 8\ ({\rm systematic})\ \ho$.

The variability of \rxj\ is sufficiently strong and erratic that the prospects 
for refining the time delay to better than $\pm 5$~percent are very good. 
Measurements of the time delays between A1, A2, and A3 also appear within 
reach from intensive optical or X-ray monitoring \citep{chartas01}. Moreover, 
further mapping of the cluster potential towards \rxj\ from \chandra, 
\xmm, \hst, and VLT observations will help bring down the systematic 
model uncertainties by determining the cluster convergence at the location
of the QSO images. Finally, the fairly high redshifts of the lens and
source result in a sensitivity of the order of 10 percent to the
adopted world model. With a smaller systematic uncertainty in the
model and independent constraints on $H_0$ the system may be used
to contrain $\Omega_m$ and $\Lambda$.  Thus, \rxj\ appears as one of 
the most useful individual lens systems for cosmological parameter 
determination and studies of the mass distribution in galaxies and clusters.

\acknowledgements

We thank the NOT Director Vilppu Piirola for granting us observing time for 
this project on a flexible basis. We are especially grateful to the many 
visiting observers at NOT who have contributed to this project by performing 
the scheduled observations. This project was conceived in 1997 while JH, AOJ, 
JPK, and JP were visiting scientists at the Center for Advanced Study in Oslo. 
We thank Rolf Stabell and Sjur Refsdal for inviting us there and for their 
kind hospitality. JH appreciates the hospitality of the OMP where some of this 
work was conducted. The project was supported by the Danish Natural Science 
Research Council (SNF). IB was supported in part by contract  ARC94/99--178 
``Action de Recherche Concert\'ee de la Communaut\'e Fran\c{c}aise (Belgium)'' 
and  P\^ole d'Attraction Interuniversitaire, P4/05 \protect{(SSTC, Belgium)}.
PLS acknowledges the support of US NSF grant AST--9616866. JP acknowledges 
partial support of grant 4697 of the Estonian Science Foundation. FPP 
acknowledges financial support by the Theoretical Astrophysics Center (TAC), 
a collaboration between the Universities of Copenhagen and Aarhus funded by 
the Danish National Research Foundation.

\clearpage

\begin{deluxetable}{crrr}
\tablecaption{Coordinates and $I$-band magnitudes for three reference 
stars in the field surrounding \rxja.
\label{refstars}}
\tablewidth{0pt}
\tablehead{
\colhead{} & \colhead{RA(J2000)} & \colhead{Dec(J2000)}  & \colhead{$I$}\\
\colhead{} & \colhead{} & \colhead{}  & \colhead{(mag)}
}
\startdata
R1       &  09:11:21.16  & +05:50:43.2 & 17.34$\pm$0.01 \\
R2       &  09:11:26.54  & +05:51:43.7 & 16.30$\pm$0.02 \\
R3       &  09:11:29.08  & +05:49:30.0 & 17.07$\pm$0.01 \\
\enddata
\end{deluxetable}

\begin{deluxetable}{cll}
\tablecaption{Estimated time delays and $I$-band magnitude differences 
for \rxja\ calculated with three different methods 
\citep{pelt96,burud01}.
\label{methods}}
\tablewidth{0pt}
\tablehead{
\colhead{} & \colhead{$\Delta$t}   & \colhead{$\Delta m$}\\
\colhead{} & \colhead{(days)}   & \colhead{(mag)}
}
\startdata
Minimum dispersion &   153$\pm$3 & \nodata  \\
$\chi^2$ fit&  147$\pm$4 & $1.88\pm0.01$\\
Iterative fit &  146$\pm$4  & 1.95--2.05 \\
\enddata
\tablecomments{The quoted uncertainties in the time delay
estimates are $1\sigma$ errors.}

\end{deluxetable}

\begin{deluxetable}{ll}
\tablecaption{Estimated time delays for \rxja\ calculated using the 
iterative method \citep{burud01}.
\label{timed}}
\tablewidth{0pt}
\tablehead{
\colhead{} & \colhead{$\Delta$t}  \\
\colhead{} & \colhead{(days)} 
}
\startdata
B--A  &   146$\pm$4  \\
B--A1 &   143$\pm$6  \\
B--A2 &   149$\pm$8  \\
B--A3 &   154$\pm$16  \\
\enddata
\tablecomments{The quoted uncertainties in the time delay
estimates are $1\sigma$ errors.}

\end{deluxetable}

\clearpage

\begin{figure}
\plotone{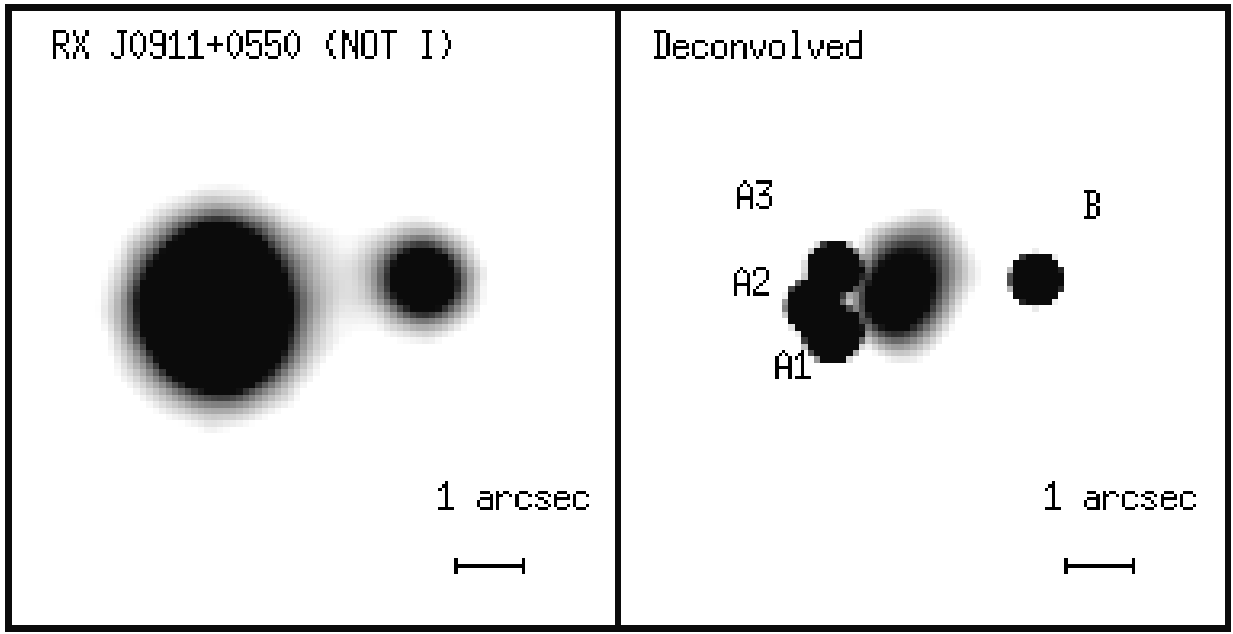}
\caption{Stacked $I$-band images of \rxja\ from a total of
$\sim$ 7 hours of exposure. The left panel shows the combined image 
(seeing $=$ 1\farcs20). The right panel shows the deconvolved  
image (FWHM = 0\farcs28). The main lensing galaxy can be seen close
to the three A components. 
North is up and East is to the left. Each frame measures 
$12\times 12$ arcsec.
\label{decima}
}
\end{figure}

\begin{figure}
\plotone{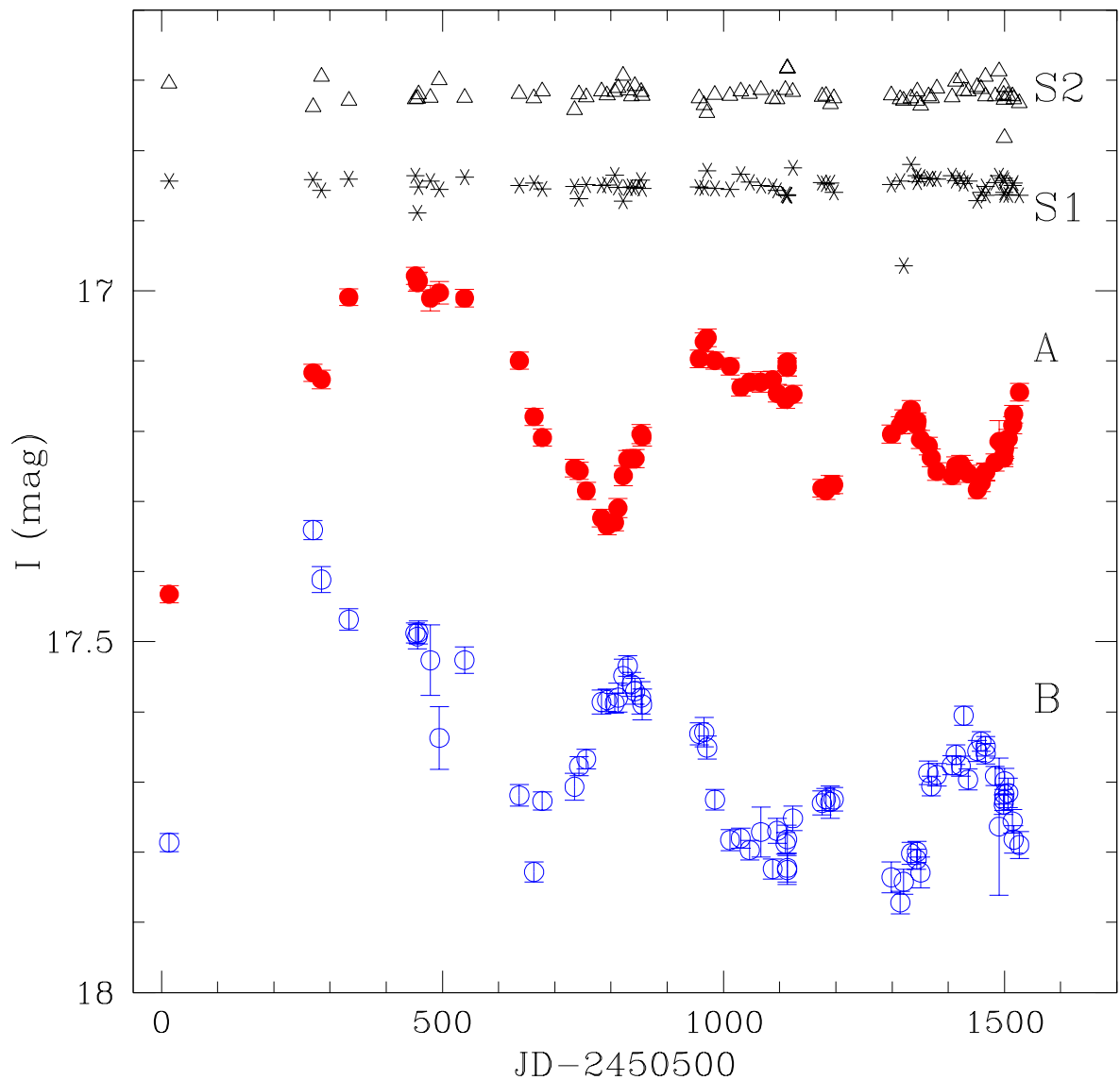}
\caption{
$I$-band lightcurves for \rxja\ A,B and two of the stars in the 
field around \rxja\ that were not used to construct the PSF. The magnitudes 
are calibrated using reference stars in the field (Table~1).
The plotted B lightcurve is offset by $-1.5$~mag and S1 is offset by 
$-0.2$~mag. The error bars represent the combined photon noise and PSF 
errors, determined as described by \citet{burud00}.
\label{lightcurve}
} 
\end{figure}

\begin{figure}
\plotone{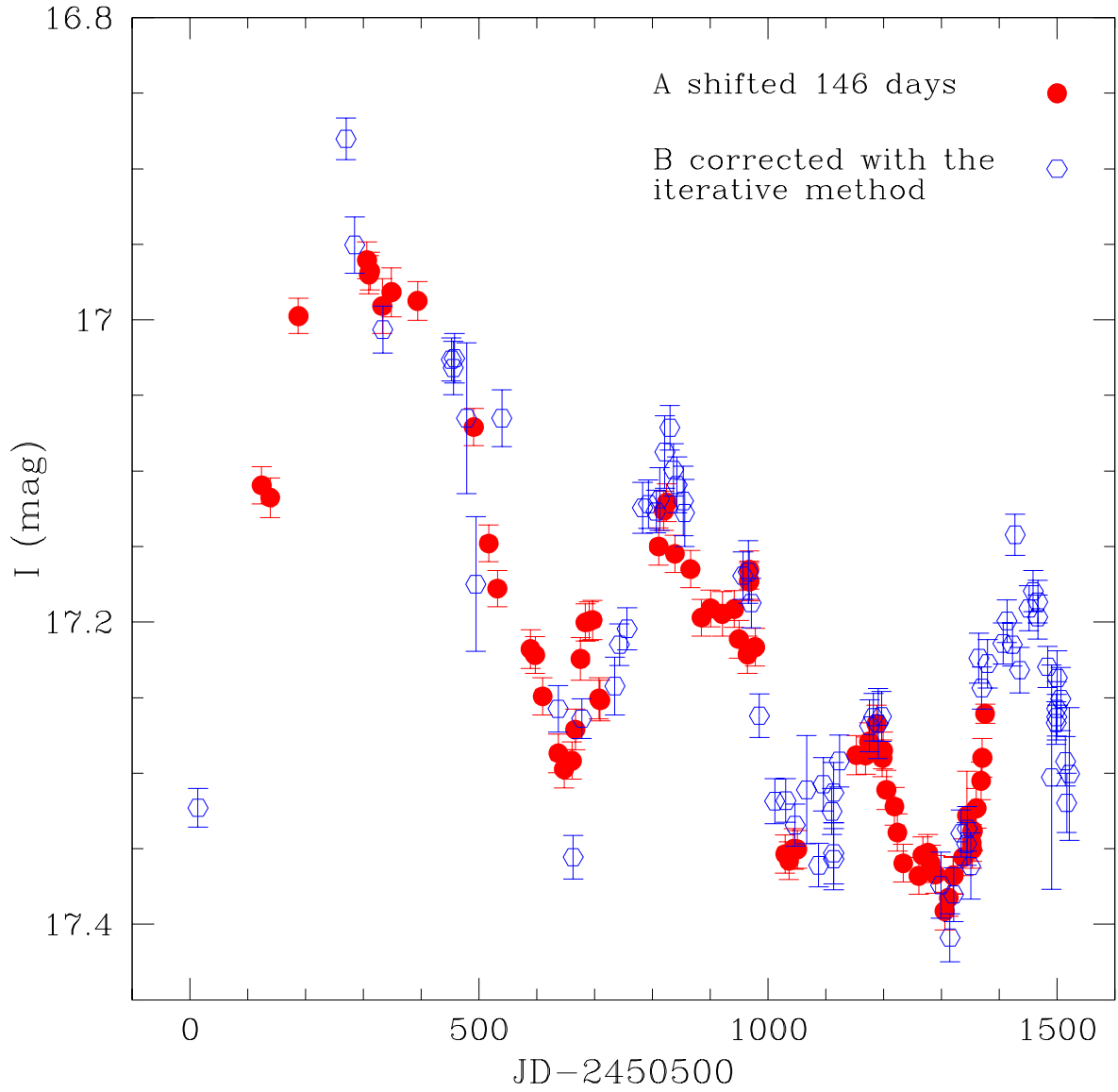}
\caption{
Combined lightcurve from both components of \rxja.
The curve of the A component is shifted forward in time by 146 days.
The B curve has been offset by values between $-1.95$ mag and $-2.05$ mag.
\label{shift}
}
\end{figure}

\end{document}